\documentclass[modern]{aastex701}

\usepackage{amssymb}
\usepackage{lipsum}
\usepackage{bm}
\usepackage{amsmath}
\usepackage{wasysym}
\usepackage{cancel}
\usepackage{xspace}
\usepackage{comment}

\newcommand{\pdf}{\mathrm{P}}

\newcommand{\tstart}{t_{\mathrm{start}}}
\newcommand{\tstartslow}{t_{\mathrm{start},\mathrm{slow}}}
\newcommand{\zstart}{z_{\mathrm{start}}}

\newcommand{\AEarth}{A_{\oplus}}
\newcommand{\Xself}{X_{\mathrm{self}}}
\newcommand{\Xwave}{X_{\mathrm{wave}}}

\begin{document}


\title{THE COSMOLOGICAL HART-TIPLER CONJECTURE}

\author[orcid=0000-0002-4365-7366,sname='Kipping']{David Kipping}
\affiliation{Columbia University, 550 W 120th Street, New York NY 10027, USA}
\email[show]{dkipping@astro.columbia.edu}

\begin{abstract}
Self-reproducing automata, so-called von Neumann machines, have been repeatedly estimated to be capable of traversing the Galaxy many times given its age. Our mere existence thus seems to exclude an aggressive variant of such a probe having ever been launched in the Milky Way. The Hart-Tipler conjecture considers this to represent contra-positive evidence to the hypothesis that other extra-terrestrial technological entities have emerged in our galaxy. Recently, several authors have extended interstellar colonization calculations to cosmological volumes, but these models are loaded with specific assumptions about behavior and emergence times. Here, we present a bare-bones model of generic artificial infections (such as but not limited to von Neumann probes) at cosmological scale in order to maximize interpretability, an approach closer to the original spirit of the Hart-Tipler calculations. Our model has just three parameters, a spontaneous spawn rate, a propagation speed ($u$) and a start time for the calculation. Accounting for cosmological expansion, we find that half the Universe is infected by today for $u=0.1c$ propagation starting 4.5\,Gyr after the Big Bang if the spawn rate exceeds approximately once per million galaxies. For near-c propagation, this becomes a billion galaxies. Over 99.9\% of cosmological volumes are filled when $u=0.1c$ if even 1 in ${\sim}10^16$ stars have previously spawned an infection. The ``cosmological Hart-Tipler'' problem therefore offers a remarkably sharp minimal-model constraint on the prevalence of aggressive, self-propagating technological behavior. We explore its implications, such as how anthropic reasoning implies such infections occur and its fine-tuning nature.
\end{abstract}



\textcolor{white}{blank}\\
\textcolor{white}{blank}\\
\textcolor{white}{blank}\\
\textcolor{white}{blank}

\section{Introduction}
\label{sec:intro}

The concept of a machine which can self-replicate has been with us many decades already \citep{freitas:1980}, conceived at least as far back as the 1960s via the posthumous writings of \citet{neumann:1966}. Further, it has been repeatedly argued that waves of self-replicating probes could plausibly engulf the entire Galaxy on time frames far shorter than its current age \citep{bracewell:1960,hart:1975,freitas:1980}. Accordingly, both \citet{hart:1975} and later \citet{tipler:1980} reasoned that our observations, and even our mere existence, appear to undermine the hypothesis that a previous technological species ever launched such a machine - sometimes dubbed ``Fact A'' (also see \citealt{brin:1983}). The Hart-Tipler conjecture is that this behaviour has never previously occurred in our galaxy, which in turn implies that technologically advanced alien species have never previously emerged in our galaxy.

The ``Hart-Tipler conjecture'' is experiencing a re-kindling of interest in recent years thanks to developments in 3D printing, artificial intelligence and commercial space flight, with \citet{ellery:2022} arguing that they are now ``imminent'' (also see \citealt{ellery:2016} and \citealt{borgue:2021}). However, it has also generated considerable debate over the decades, most famously via the pushback of \citet{sagan:1983} (the ``Sagan-Tipler debate''). \citet{sagan:1983} offered a host of critiques; for example, that ``our Solar System may have been intentionally bypassed because of the evolution of intelligence here'' - what one might reasonably classify as a manifestation of the Zoo Hypothesis \citep{ball:1973}. However, the Zoo Hypothesis requires a cultural hegemony across galactic scales that has been argued to be somewhat contrived due to the sheer scale of the Galaxy and the finite speed of light \citep{forgan:2017}. \citet{sagan:1983} suggested another resolution to be that some civilizations built anti Von Neumann machines; probes that prey on their counterpart (see also \citealt{stull:1979}). But, again, this suggestion does not stand up to more detailed scrutiny, with \citet{forgan:2019} finding that, in general, this does not meaningfully reduce the probe population. 

A third line of inquiry and skepticism is that the simple toy model of galactic colonization utilized by \citet{hart:1975} and \citet{tipler:1980} misses crucial factors that impede expansion waves. For example, \citet{ashworth:2014} consider range limits on probes, \citet{jones:1976} argued that colonization explosion could be blunted by population control and \citet{gros:2005} suggested cultural forces might explain their absence. However, \citet{wright:2014} have argued that any solution to the broader Fermi Paradox that requires universal choices or behaviour is highly dubious, termed the ``monocultural fallacy'' (see also \citealt{hart:1975}).

A related solution to the Hart-Tipler conjecture is that we live in a local void, which again harkens back to \citet{sagan:1983} who suggested that expansion would more realistically follow a nonlinear diffusion process. In particular, \citet{landis:1998} suggested percolation theory could produce large voids which naturally explain Fact A, and ever more sophisticated 2D/3D simulations of galactic colonization have studied this further (e.g. see \citealt{bjork:2007,cotta:2009,forgan:2009,hair:2013}), even including 3D stellar thermal motion and galactic shear \citep{nellen:2019}.

From the above, we can see that a common theme is that one can potentially resolve SETI optimism with Fact A by appending additional parameters and complexity to the model of how von Neumann probes spread across the stars. But, of course, this is not a unique property of the Hart-Tipler problem, it's a general statistical feature - complex models maximize explanation, whilst simpler models maximize prediction \citep{shmueli:2011}. Further, we highlight that one should delineate between the notion of a biological species gradually settling inviting worlds (and the sociological and ecological complexities that may entail) versus the indifferent, machine-like expansion of von Neumann probes. For example, \citet{nellen:2019} suggest that the ``Aurora effect'' plays a significant role in producing unpopulated voids, the fact that some worlds will be more habitable than others, but this is plausibly muted for machine replication.

Ultimately, we argue that the utility of a highly simplified model, such as that originally used by \citet{tipler:1980}, is historically self-evident. Whether it be due to a lack of aliens or missing model complexity, Fact A \textit{demands} explanation, a statement clearly borne out by the rich literature on this topic.

Almost all studies debating the Hart-Tipler conjecture focus on our galaxy. But, if we take seriously the proposition that probes could traverse it in cosmically short windows, then we must also relax the implicit assumption of island universes. Even in the original work of \citet{tipler:1980}, it was suggested that most of the entire Universe should be filled by now, leading to Tipler's provocative title ``Extraterrestrial intelligent beings do not exist''. However, neither \citet{tipler:1980} nor his subsequent dialogue \citep{tipler:1982,tipler:1983} ever wrote down a formal calculation accounting for cosmological expansion, which acts as a crucial drag to sub-light expansion waves.

The first appearance of such a calculation was offered in \citet{olson:2015}, who framed the model in terms entities that convert a parameterized fraction of matter into radiation at a parameterized rate but with a parameterized delay time. \citet{hanson:2021} offered another calculation but predicated their approach upon the ``Hard Steps'' model of evolution \citet{carter:2008} - an idea which has been forcefully challenged \citep{mills:2025}. We also highlight the work of \citet{armstrong:2013}, which considers the energetics of inter-galactic expansion rather than a cosmological model of it.

As noted above, the published cosmological variants of the Hart-Tipler conjecture have leap-frogged the conventional approach of first introducing a simplified, baseline model and instead immediately proposed models with bespoke complexities offering a greater surface area of potential critique (e.g. \citealt{mills:2025}). Although these investigations are certainly valuable, recall that we argued earlier that the power of the original Hart-Tipler argument was in its simplicity. After all, all models are wrong, but some are useful \citep{box:1976}. In the spirit of the original Hart-Tipler conjecture, then, we here propose the simplest (most ``economic'') cosmological extension one can reasonably get away with - a model parameterized by a spawn rate, a propagation speed and a starting cut-off time for probes to emerge.

\section{A Simple Model of Cosmological Epidemics}
\label{sec:model}

\subsection{Infection model}

Conceptually, our model works as follows. First, since we are principally interested in cosmological scales and average behavior, we assume a statistically homogeneous population of galaxies in an FLRW universe with scale factor $a(t)$. Consider that the galaxies have conserved comoving number density $n_G$, such that $n_G\,\mathrm{d}V_c$ is the expected number of galaxies in a comoving volume element $\mathrm{d}V_c$. In our later numerical evaluations in Section~\ref{sec:implications}, we adopt $n_G=10^{-2}\,{\rm cMpc}^{-3}$, a conventional order-of-magnitude comoving number density. Since the results are controlled by the product $n_G\lambda$, alternative choices of $n_G$ simply rescale the inferred per-galaxy spawn rate.

Our model assumes that each galaxy can transition from an uninfected state (U) to an infected state (I) via two pathways, either of which constitute a permanent state change. The first method is that a galaxy can spontaneously mutate from an uninfected state (U) to an infected state (I) with constant hazard rate of $\lambda(=1/\tau)$ per galaxy per unit cosmic time. This spontaneous self-mutation rate is more conveniently described as simply the spawn rate. For simplicity, we treat this state change as instantaneous on cosmological timescales, justified by the fact that intra-galactic light crossing times are much shorter than the total time under consideration.

The second pathway for a state change is by external infection: once a galaxy becomes infected, it immediately emits a spherical infection front that propagates at physical peculiar speed $u$ relative to the local Hubble flow. Any galaxy swept over by the front becomes infected. Unlike \citet{olson:2015}, our model makes no assumptions that infected regions would present specific observable signatures, but rather the principal effect of infection is that it nullifies said region's habitability. The relevant assumption is not literal sterilization, but that transformed regions cease to contribute observers in our reference class.

Note that our framing avoids the explicit assumption of von Neumann probes, but rather casts the problem in the more general form of some kind of artificial infection. We use the term ``infection'' in a mathematical sense only: a self-propagating transition from a habitable/untransformed state to an uninhabitable or observer-suppressing state. No biological analogy is intended.

The infection fronts are mathematically modeled as spherical wave fronts, which can be interpreted either as literal isotropic expansion or as an effective envelope for a sufficiently dense directed-probe strategy (e.g. \citealt{crick:1973}). In this way, the model could be considered to encompass a variety of infection modes. Indeed, our intention here is to avoid conditioning the model upon a specific mechanism because any assumptions of ``advanced'' behavior often age poorly (e.g. Martian canals; \citealt{chambers:1999}), since we cannot reliably predict what new technological paradigms might arise.

The goal of this work is now to compute the expected fraction of galaxies that are infected ($f$) at cosmic time $t$, and as a function of the mutation rate $\lambda$ and infection speed, $u$.

\subsection{Self-mutation probability for a single galaxy}

First, consider one particular target galaxy, ignoring infection waves. Since its spontaneous mutation rate is constant ($\lambda$), then the probability that it has \textit{not} mutated during the interval $[\tstart,t]$ is the zero-event probability of a Poisson process:

\begin{align}
\pdf_{\rm no\,self}(t) &= \exp[ -\Xself(t) ],
\label{eqn:Pnoself}
\end{align}

where we have defined

\begin{align}
\Xself(t)
\equiv& \int_{\tstart}^t \lambda\,dt' = \lambda (t - \tstart).
\label{eqn:Xself}
\end{align}

\subsection{Comoving radius of an infection front}

Next, consider an infection seed that appears at cosmic time $t'$. It emits a spherical infection wave moving at physical peculiar speed $u$. In an FLRW universe, a radial physical distance interval is related to a comoving interval by

\begin{align}
\mathrm{d}r_{\mathrm{phys}} &= a(t)\,\mathrm{d}\chi,
\end{align}

where $\chi$ is comoving distance. Therefore, an infection front moving with physical speed $u$ obeys

\begin{align}
\frac{ \mathrm{d}\chi }{ \mathrm{d}t } &= \frac{ u }{ a(t) }.
\end{align}

Hence, the comoving radius reached by a wave born at time $t'$ and observed at time $t$ is

\begin{align}
\chi(t,t') &=
\int_{t'}^{t} \frac{ u\,\mathrm{d}t'' }{ a(t'') }.
\label{eqn:chi}
\end{align}

\subsection{The infection volume relevant to a target galaxy}

To determine whether a chosen target galaxy has been infected by a time $t$, we ask whether any infection seed appeared early enough and close enough for its infection wave to have reached the target. A seed born at time $t'$ can infect the target by time $t$ if it lies within comoving radius $\chi(t,t')$ of the target. Therefore, the relevant comoving volume around the target is

\begin{align}
V_{\mathrm{reach}}(t,t') &=
\frac{ 4 \pi }{ 3 } \chi(t,t')^3.
\end{align}

Substituting $\chi$ using Equation~(\ref{eqn:chi}), one finds

\begin{align}
V_{\rm reach}(t,t') &=
\frac{ 4 \pi }{ 3 }
\Big[ \int_{t'}^{t} \frac{ u\,dt'' }{ a(t'') } \Big]^3.
\label{eqn:Vreach}
\end{align}

Note that this volume is not a physical volume at time $t'$; it is the comoving volume containing all possible seed galaxies whose waves would reach the target by time $t$.

\subsection{Rate of external infection opportunities}

The expected number of galaxies in a comoving volume $\mathrm{d}V_c$ is $n_G\,\mathrm{d}V_c$, and each such galaxy has a spawn rate of $\lambda$. Therefore, the rate at which infection seeds appear per unit comoving volume is

\begin{align}
\Gamma_c &= n_G \lambda.
\label{eqn:gamma}
\end{align}

During an interval $\mathrm{d}t'$, the expected number of spontaneous infection seeds appearing inside the reach volume $V_{\mathrm{reach}}(t,t')$ is

\begin{align}
\mathrm{d}X_{\rm wave}(t;t') &=
\Gamma_c\,V_{\rm reach}(t,t')\,\mathrm{d}t'\nonumber\\
\quad&= \frac{ 4 \pi \lambda n_G }{ 3 }
\Big[ \int_{t'}^{t} \frac{ u\,\mathrm{d}t'' }{ a(t'') } \Big]^3 \mathrm{d}t'.
\end{align}

where the second line substitutes Equations~(\ref{eqn:gamma}) \& (\ref{eqn:Vreach}). Integrating over all possible seed times gives the expected number of external infection opportunities:

\begin{align}
\Xwave(t) &=
\int_{\tstart}^t \mathrm{d}\Xwave(t;t'),\nonumber\\
\qquad&= \frac{ 4 \pi \lambda n_G }{ 3 }
\int_{\tstart}^{t} \Big[ \int_{t'}^{t} \frac{ u\,\mathrm{d}t'' }{ a(t'') } \Big]^3 \mathrm{d}t'
\end{align}

Equivalently, since $u$ is assumed to be a constant, then

\begin{align}
\Xwave(t) &=
\frac{ 4 \pi \lambda n_G u^3 }{ 3 }
\int_{\tstart}^{t} \Big[ \int_{t'}^{t} \frac{ \mathrm{d}t'' }{ a(t'') } \Big]^3 \mathrm{d}t'
\label{eqn:Xwave}
\end{align}

Equation~(\ref{eqn:Xwave}) now yields the expected number of independent external infection events that would have reached the target galaxy by time $t$, before accounting for the fact that one infection is sufficient. Note that the wave term treats other galaxies as a continuum Poisson field. The target galaxy’s own spontaneous mutation hazard is therefore added separately to recover the correct $u\to0$ limit.

\subsection{Combining self-mutation and wave infection}

A galaxy remains uninfected by time $t$ only if \textit{both} of the following conditions hold true: i) it has not spontaneously self-mutated, and ii) no external infection wave has reached it. Under the assumptions of independent Poisson processes, the probability of no self-mutation was found earlier in Equation~(\ref{eqn:Pnoself}) as $\exp[-\Xself(t)]$, and by extension the probability of no external wave infection will be $\exp[-\Xwave(t)]$. Thus, the total survival probability - the probability of being uninfected - will be

\begin{align}
\pdf_U(t) &= \exp[ -\Xself(t) ] \exp[ -\Xwave(t) ],\nonumber\\
\qquad&= \exp[ -\Xself(t) - \Xwave(t) ].
\end{align}

Since $f(t)$ is the fraction that has become infected by time $t$, it trivially follows that

\begin{align}
f(t)  &= 1 - \pdf_U(t),\nonumber\\
\qquad&= 1 - \exp\big[ -\Xself(t) - \Xwave(t) \big].
\end{align}

Substituting the two terms, one obtains

\begin{align}
f(t) &= 1 - \exp\Bigg[
-\lambda(t-\tstart)
-\Big(\frac{ 4 \pi \lambda n_G u^3 }{ 3 }\Big)
\int_{\tstart}^{t} \Big[ \int_{t'}^{t} \frac{ \mathrm{d}t'' }{ a(t'') } \Big]^3 \mathrm{d}t'
\Bigg].
\label{eqn:fI}
\end{align}

Because newly infected galaxies emit fronts with the same speed and zero delay, secondary fronts cannot outrun the envelope generated by the original seed in a homogeneous background. The wave term therefore need only integrate over spontaneous seeds; the subsequent cascade is already represented by the spherical growth volume.

\subsection{Present-day expression in redshift}

For cosmological applications it is convenient to express $f$ in terms of redshift, $z$. In this framing, today represents $z=0$ and the scale factor behaves as $a = 1/(z+1)$. Accordingly, the relation between cosmic time and redshift is

\begin{align}
\mathrm{d}t &= -\frac{ \mathrm{d}z }{ (1+z) H(z) },
\end{align}

where $H(z)$ is the Hubble parameter as a function of redshift. The elapsed cosmic time between $\zstart$ and today is therefore

\begin{align}
\Delta t &= \int_{\tstart}^{t_0} \mathrm{d}t,\nonumber\\
\qquad   &= \int_0^{\zstart}     \frac{ \mathrm{d}z }{ (1+z) H(z) }.
\end{align}

Accordingly, it follows from Equation~(\ref{eqn:Xself}) that

\begin{align}
\Xself(0) &=
\lambda \int_0^{\zstart} \frac{ \mathrm{d}z }{ (1+z) H(z) }.
\label{eqn:Xselfnow}
\end{align}

We can now apply the same approach to the infection wave. Consider the comoving radius reached today by a wave born at redshift $z'$. Since

\begin{align}
\frac{ \mathrm{d}t }{ a(t) } &= -\frac{ \mathrm{d}z }{ H(z) },
\end{align}

then we have

\begin{align}
\chi(z') &= \int_{ t(z') }^{t_0} \frac{ u\,\mathrm{d}t }{ a(t) },\nonumber\\
\qquad   &= \int_0^{z'}          \frac{ u\,\mathrm{d}z }{ H(z) }.
\end{align}

The last ingredient we need is that seed-time element becomes

\begin{align}
\mathrm{d}t' &= -\frac{ \mathrm{d}z' }{ (1+z') H(z') }.
\end{align}

Combining these results with Equation~(\ref{eqn:Xwave}), we may now reverse the integration limits from early-to-late time into redshift from $0$ to $\zstart$, giving us the external-wave term today as

\begin{align}
\Xwave(0) &=
\frac{ 4 \pi \lambda n_G u^3 }{ 3 }
\int_0^{\zstart}
\Big[ \int_0^{z'} \frac{ \mathrm{d}z }{ H(z) } \Big]^3
\frac{ \mathrm{d}z' }{ (1+z') H(z') }.
\label{eqn:Xwavenow}
\end{align}

Therefore, the present-day infected fraction is

\begin{align}
f(0) &= 1 - \exp\Big[ -\Xself(0) - \Xwave(0) \Big].
\end{align}

Explicitly,

\begin{align}
f(0) &= 1 - \exp\Bigg[
-\lambda \int_0^{\zstart} \frac{\mathrm{d}z}{(1+z)H(z)}
- \frac{ 4 \pi \lambda n_G u^3 }{ 3 }
\int_0^{\zstart}
\Big[ \int_0^{z'} \frac{ \mathrm{d}z }{ H(z) } \Big]^3
\frac{ \mathrm{d}z' }{ (1+z') H(z') }
\Bigg].
\label{eqn:fInow}
\end{align}

Finally, to solve the above, we need a cosmological model. Assuming a flat $\Lambda$CDM Universe, one may write

\begin{align}
H(z) &\simeq H_0 \sqrt{ \Omega_m (1+z)^3 + \Omega_{\Lambda} },
\label{eqn:Hz}
\end{align}

where we have ignored the influence of radiation, $\Omega_r$, given that it contributes negligibly over the low-redshift range relevant to our fiducial calculations. Substituting $H(z)$ into Equation~(\ref{eqn:fInow}) gives a fully specified numerical model for $f(0)$.

\section{Implications}
\label{sec:implications}

\subsection{Choosing a Cut-off Time}

Using Equation~(\ref{eqn:fInow}), one may numerically evaluate the current fraction of the Universe that has become infected, $f(0)$, for any choice of spawn rate, $\lambda$, propagation speed, $u$, and some choice for the cutoff redshift before which probes are prohibited, $z_{\mathrm{start}}$. This latter parameter is distinct from the other two in that we can put some constraints on its value based on astronomical observations, which we explore here.

One might consider that the earliest point by which a technological civilization could plausibly emerge to be set by 1) the earliest formation of rocky planets\footnote{Of course, one could be less conservative and assume non-terrestrial planets are valid seats for life too, but we avoid such speculation here.}, and, 2) sufficient time for life to develop and evolve to sufficiently complex creatures capable of triggering infection waves.

The latter of these is perhaps easiest to discuss since we have a singular example to govern our inference - ourselves. The Earth's current age is $\AEarth = (4.54\pm0.05)$\,Gyr \citep{dalrymple:2001} and thus it, at least here, it took that long to go from planet formation to the Anthropocene. Provided that the leap from modern technology to that required for triggering infections is ${\ll}$Gyr, $\AEarth$ therefore sets a conservative lower bound on the evolutionary timescale.

If technological species can only develop on rocky worlds, then certainly such bodies will not present at $z=\infty$ - there must be some delay in order for metals to cook in the bowels of the first stars and then subsequently disperse and coalesce into planets \citep{johnson:2012}. The correlation between planet occurrence and host star metallically is unequivocal for giant planets \citep{fischer:2005,johnson:2010,thorngren:2016}, but for rocky planets, the correlation is far weaker and its strength remains debated \citep{wang:2015,buchhave:2015,boley:2024}. In the core-accretion framework, this implies that massive $(\gtrsim 10\,M_{\oplus})$ rocky bodies, the presumed embryos of giant planets, would be rare in the early Universe, but the formation of Earth-mass worlds appears less impinged. Indeed, despite the challenges of precisely aging stars, numerous exoplanets of likely rocky composition have been discovered around ancient thick-disk stars. For example,
Kepler-10\,b is a ${\sim}1.4$\,$R_{\oplus}$ planet orbiting a $(11.9\pm4.5)$\,Gyr host \citep{batalha:2011} and
TOI-561\,b is a ${\sim}1.5$\,$R_{\oplus}$ planet orbiting a $(10\pm3)$\,Gyr star \citep{weiss:2021}. Perhaps of greatest relevance is Kepler-444, which hosts five sub-Earth sized planets orbiting a $(11.2\pm1.0)$\,Gyr star \citep{campante:2015} - supported by  a follow-up study by \citet{buldgen:2019} who obtain $(11.0\pm0.8)$\,Gyr. The discoverers, \citet{campante:2015}, remark that Kepler-444 ``implies that small, including Earth-size, planets may have readily formed at earlier epochs in the Universe’s history when metals were more scarce''. Accordingly, current observations support that rocky planets had formed in the Universe by 11\,Gya, when it was 2.8 Gyr old.

Together, then, the slowest and most conservative choice for $\tstart$ appears to be ${\sim}2.8$\,Gyr plus $\AEarth\sim4.5$\,Gyr, yielding $\tstart = \tstartslow \equiv 7.3$\,Gyr, or 6.5\,Gya (redshift $0.72$). Paired with this strictly conservative estimate, we consider it useful to also propose a less stringent estimate. After all, it would be rather surprising if Kepler-444 turned out to be oldest rocky planet system in the Universe and that we represent the absolute fastest example of evolution to have ever transpired. Surely, these processes must in some cases occur even sooner. We thus adopt $\tstart = \AEarth = 4.5$\,Gyr ($\zstart = 1.44$) as a second fiducial choice, on the basis that there are no obvious barriers to rocky planet formation yet identified.

We remind the reader that it's perfectly reasonable to challenge what the precise correct fiducial parameter set should be. The purpose here is merely to offer a set of numbers which allows us to make numerical progress yet are at least plausible.

\subsection{It's Easy to Eat a Universe}

The most startling finding of this work is just how easy it is for infection waves to have filled the cosmos by now. We demonstrate this by first highlighting some simple example calculations and then we'll offer deeper parameter explorations. In what follows, we consider two fiducial choices for propagation speed, first a ``slow'' wave of $u=0.1c$, which is both consistent with the original work of \citet{tipler:1980}, as well as being plausibly within reach for humanity within a few decades (e.g. Breakthrough Starshot; \citealt{starshot:2018}). Our alternative limit, a ``fast'' limit, is $c$ itself, or really probes which asymptotically approach $c$.

Using Equation~(\ref{eqn:fInow}), we can iteratively solve for the spawn rate $\lambda$ which satisfies $f=f_{1/2} = 0.5$ - the point at which half of the cosmos has been filled. Since filling tends to occur as a rapid and broad phase transition, $f_{1/2}$ represents as a crucial tipping point.

The most conservative combination of fiducial parameters (which in turn allows for the most optimistic spawn rates) are slow probes ($u=0.1c$) and $t = \tstartslow = 7.3$\,Gyr, for which $f=f_{1/2}$ occurs when $\lambda_{1/2} = 6.5 \times 10^{-7}$ per galaxy per Gyr. Thus, the mean fraction of galaxies to have ever spawned an infection would be $\lambda_{1/2} (t_0-\tstartslow) = 4.2 \times 10^{-6}$, which is 1 in 240,000 galaxies, or equivalently 1 in 24 quadrillion stars\footnote{See the Appendix A for the conversion from galaxies to stars.}

Adjusting $\tstart$ to $\AEarth$, which recall we argue is a more realistic cutoff time, but sticking with slow probes, yields $\lambda_{1/2} = 1.1 \times 10^{-7}$ per galaxy per Gyr. This corresponds to an infection spawning in 1 in $(1.0 \times 10^6)$ galaxies, or 1 in 100 quadrillion stars.

These two calculations already illustrate the profound challenge that the cosmological Hart-Tipler problem poses. If we switch to fast probes with speeds that asymptotically approach $c$, the crisis is greatly exacerbated by three orders-of-magnitude. In that case, half of the Universe is filled if a mere 1 in a billion galaxies have ever spawned an infection (using $\tstart=\AEarth$), or 1 in $10^{20}$ (100 quintillion) stars.

\begin{figure*}
\centering
\includegraphics[width=17.0 cm]{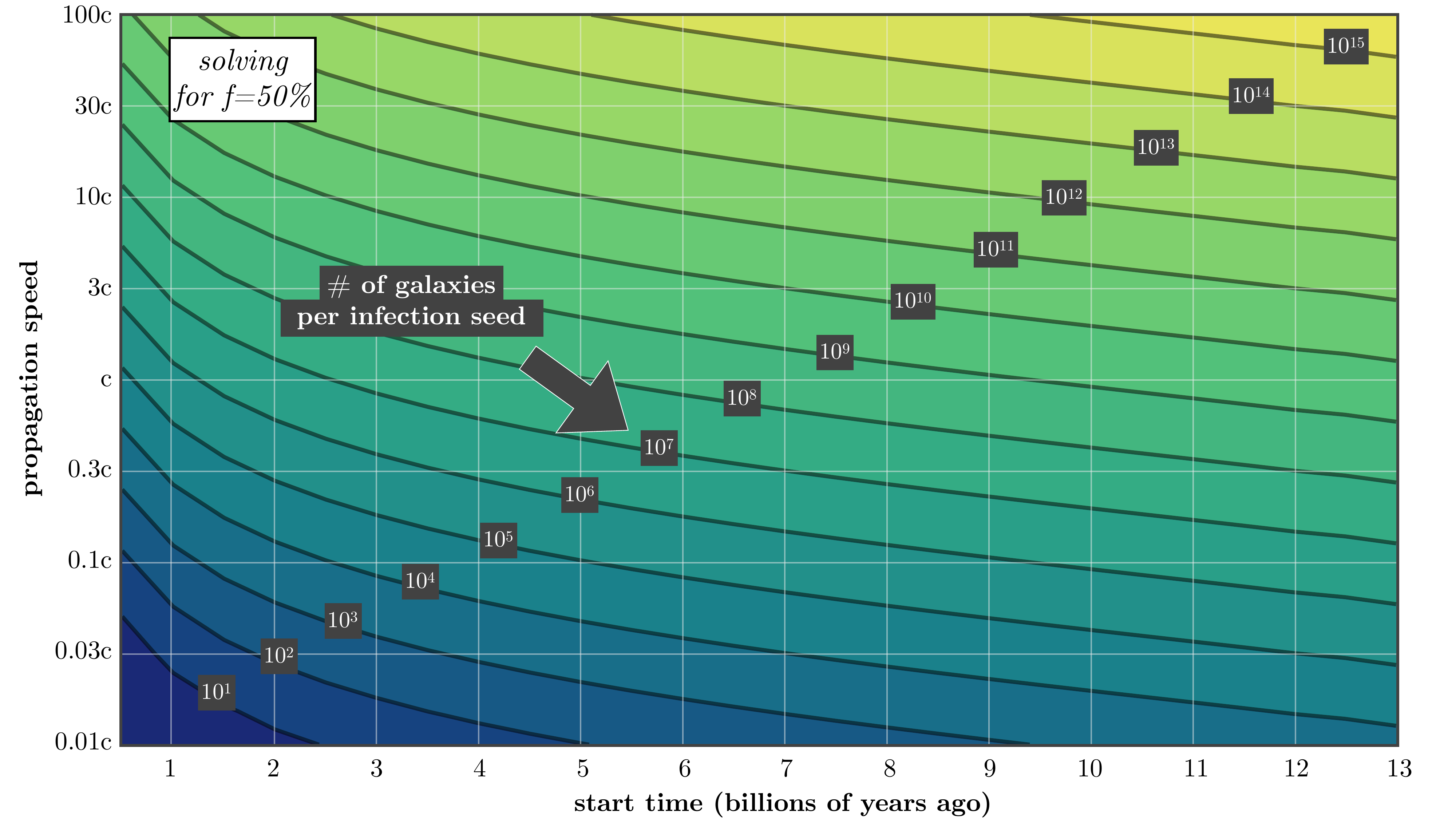}
\caption{\emph{
A grid of solutions that produce a cosmos precisely half-filled by an infection that has some spontaneous spawn rate within galaxies and then emanates an infection wavefront propagating at a speed given by the $y$-axis. The $x$-axis varies the earliest time for which we allow infection seeds to spawn. The contours denote the solved spawn rate to produce half-filling, framed in terms of the mean number of galaxies required to produce one infection seed.
}
} 
\label{fig:fhalf}
\end{figure*}

Figure~\ref{fig:fhalf} presents solutions for $f_{1/2}$ across a wide grid of possible input parameters. Since some authors have proposed relativistic workarounds for superluminal flight, such as that of \citet{alcubierre:1994}, for completeness we include $u>c$ speeds as well, although we don't comment on these further.

\subsection{Could We Live in the Void?}

\begin{figure*}
\centering
\includegraphics[width=17.0 cm]{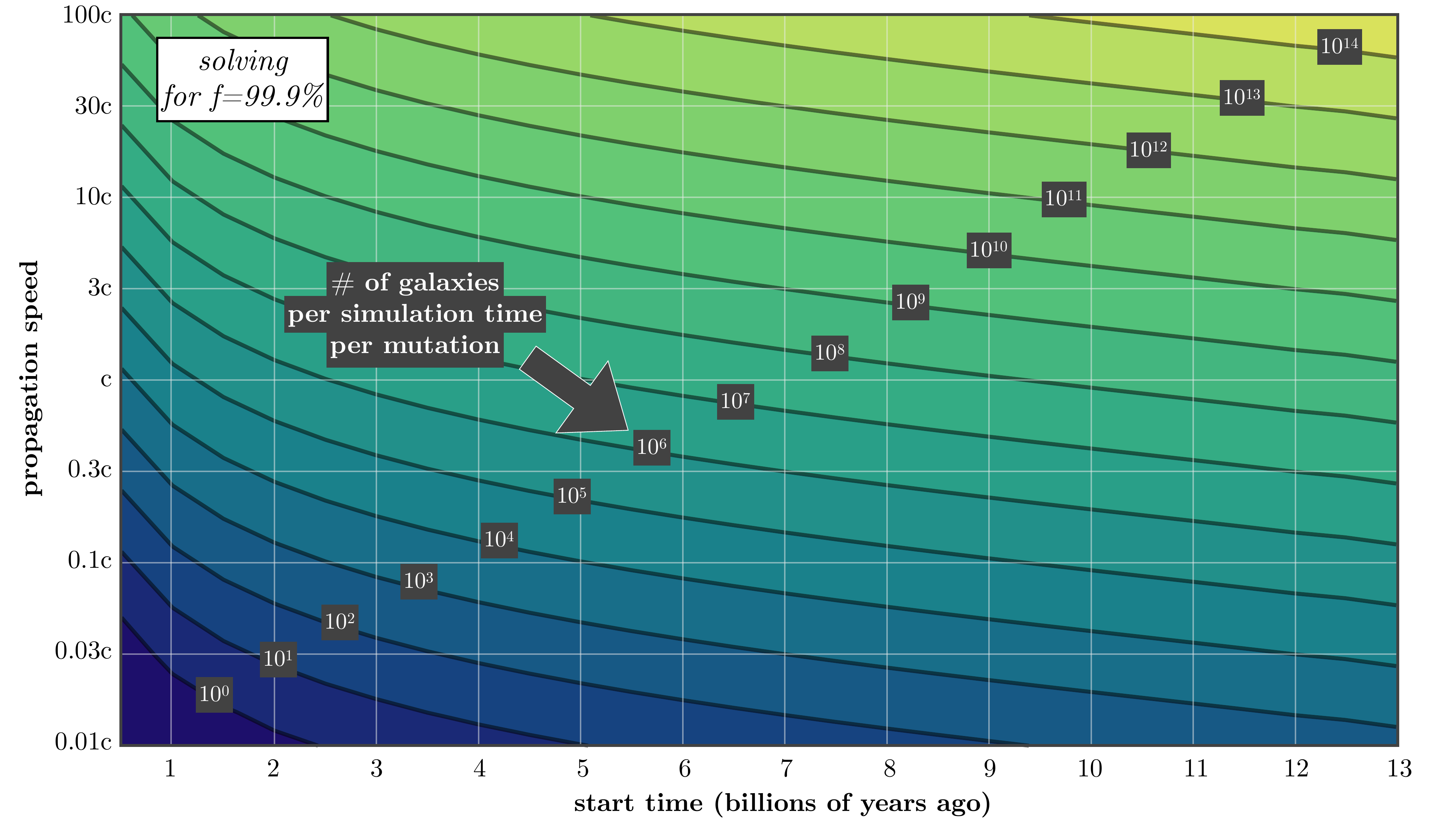}
\caption{\emph{
Same as Figure~\ref{fig:fhalf} but solving for $f=0.999$.
}
} 
\label{fig:f999}
\end{figure*}

One might argue that any scenario for which half the Universe is filled poses no logical contradiction to our existence. We would simply live in the other half. We remind the reader though that $f_{1/2}$ represents a tipping point of a rapid phase transition, and even small positive perturbations to the fiducial parameters quickly fills the cosmos. To show this, we repeated the grid of calculations shown in Figure~\ref{fig:fhalf} but solving for $f=99.9$\% instead. The results, presented in Figure~\ref{fig:f999}, reveal a broadly similar set of solutions, with a modest shift in the contours in logarithmic space.

\subsection{Anthropic Reasoning}

One might still argue that even a scenario of $f=99.9$\% imparts no logical inconsistency via the weak anthropic principle \citep{carter:1974}, or more sharply using the Self-Sampling Assumption (SSA), under which we reason as though we are randomly sampled from an appropriate reference class of actually existing observers \citep{bostrom:2013}. Even if 99.999...\% of the Universe is filled, in an infinite Universe there will always be some remaining voids. These would then be the only places where habitable planets remain and thus it would not be surprising that we emerged in such a void - indeed it's necessarily so. Thus, our existence is wholly compatible with the $f\to0$ or $f\to1$ scenarios - states that we consider to be deep and saturated in the sense that each encompass vast ranges of $\lambda$ values. If the only datum conditioned upon is our existence in an uninfected region, then SSA by itself supplies little leverage against any choice of $f$, provided both scenarios contain at least some observers in the relevant reference class. If held to be true, the Hart-Tipler problem has little inferential power - at least when framed in terms of our mere existence as the conditional.

This argument can appear quite convincing, but it is predicated on the assumption that there is a single Universe and it couldn't be any other way. If, instead, we permit the existence of a multiverse, in whatever abstract terms one wishes, then there should be a landscape of filled Universes and, vice versa, ones with plenty of habitable real estate remaining. Selection bias would seem to indicate that we should emerge in the latter.

However, this too misses an important nuance. Presumably, the probability of a technological species developing is proportional to the spawn rate of artificial infections. Accordingly, universes with $f\to0$ may not be so conducive to our emergence after all, since their low spawn rate is implies that their intrinsic parameters are tuned to somehow greatly inhibit the development of complex life. This re-framing leans on what is known as the Self Indication Assumption (SIA) in anthropic reasoning \citep{bostrom:2013}.

Under SIA, hypotheses containing more observers in the relevant reference class receive greater anthropic weight, all else being equal. With such an SIA-weighting scheme, one can consider the favored value of $f$ that this implies. For fixed $u$, $\tstart$, and present cosmic time, Equation~(\ref{eqn:fInow}) has the form $f(\lambda)=1-\exp[-k\lambda]$, where $k$ is independent of $\lambda$. If we further assume that the emergence rate of observers in the relevant reference class is proportional to the same underlying process that sets $\lambda$, then the expected number of extant observers scales as

\begin{align}
N_{\mathrm{obs}}(\lambda) \propto& \lambda[1-f(\lambda)],\nonumber\\
\qquad                           &=\lambda \exp[-k\lambda].
\end{align}

This heuristic SIA weighting is maximized at $k\lambda=1$, corresponding to $f=1-e^{-1}\simeq0.63$. Although such half-filling might seem fine-tuned \citep{lewis:2025}, implying that we live in a universe balanced on the knife-edge of the infection tipping point, it naturally maximizes the number of observers. Such a scenario balances the need for a sufficiently high spawn rate such that civilizations frequently arise, with a sufficiently low rate such that plenty of habitable volume remains. Indeed, the SIA naturally produces an inverted-bowl like weighting for $f$ peaking at ${\sim}1/2$, since both $f\to0$ and $f\to1$ lead to universes largely devoid of observers such as ourselves.

The anthropic arguments made thus made have exploited the SSA and SIA (at least in their original forms), both of which focus on the number of observers. Further, the anthropic reasoning thus far has implicitly treated the current epoch as a fixed conditional, rather than worrying about why we don't live in the far future or past, for example. However, such conditioning is obviously dubious when speaking of multiverses and abstract landscapes of possible parameter sets. Indeed, more modern version of SIA consider ``observer moments'' instead \citep{elga:2000,cirkovic:2001}. Such an approach necessarily requires extending our model to future epochs, which we have thus far avoided. Further, we argue that our simple model cannot be extended into the far future without losing its utility, because the diminishing rate of star formation and finite stellar lifetimes would become major factors that we omit. This is where the more complex modeling approaches offered by \citet{olson:2015} and \citet{hanson:2021} would be required.

Nevertheless, we can at least qualitatively briefly consider the implications of weighting by observer moments. One should expect that the number of observer moments will be maximized by delaying the onset of the $f_{1/2}$ tipping point as long as possible; a time we dub $T_{1/2}$. If we treat $c$ as immutable, then the only lever to achieving this to decrease $\lambda$. As shown in Appendix B, our model implies a half-filling time scaling as $T_{1/2} \propto \lambda^{-1/2} u^{-3/2}$ for a matter dominated universe, and $T_{1/2} \propto \lambda^{-1/4} u^{-3/4}$ for a static one. In either case, the observer moment maximization corresponds to maximizing $\lambda T_{1/2}$, which combines to $\propto \lambda^x$ where $x>0$. Thus, we broadly expect this argument to favor universes with large $\lambda$, which rapidly fill. Such universes produce many observers and thus observer moments, but they are ultimately short-lived and transient - truncated by the rapid filling process. This result ends up being thematically consistent with observer-weighted result then. In a sense, this picture gives SETI some hope for there would be other out there. But, on the other hand, it would also imply this state is temporary and rapid filling is imminent. However, we once again emphasize that this scaling argument should not be interpreted as a full observer-moment calculation, since such a calculation would require finite evolutionary delay times, stellar lifetimes, star-formation history, and the temporal duration of observer-producing civilizations.

\section{Discussion}
\label{sec:discussion}

\subsection{Concession of Model Simplicity}

We acknowledge that our model is extremely simple, governed by just three parameters, $\lambda$, $u$ and $\tstart$. But this is not a bug, but rather a design feature. We certainly welcome more sophisticated treatments, such as adding additional parameters to account for probabilistic spreads, behaviours, probe mutations, etc. However, we firmly believe that complexity must first build upon a simple baseline model to make it easily interpretable. Every new parameter adds potential confusion to what drives simulation outcomes, as well representing new points of logical vulnerability. For example, a model with many parameters/assumptions could have just one that invites criticism of its appropriateness, and then it becomes easy to dismiss the entire concept due to that one small crack. Indeed, the related idea of ``Grabby Aliens'' \citep{hanson:2021} is predicated upon a particular assumption about how evolution proceeds that has been criticized \citep{mills:2025}, thereby blunting its impact. Similarly, the model of \citet{olson:2015} is more elaborate than is required for the narrower Hart-Tipler question pursued here. However, for completeness, we provide a comparison of our model to the work of \citet{olson:2015} in Appendix C.

We thus defend our decision to make this model ``as simple as it can be, but no simpler'' (to quote Einstein) on the grounds of serving as a baseline for future work, maximizing interpretability and adherence to Occam's razor.

\subsection{Why the SIA Tacitly Supports the Infection Model}

In anthropic reasoning, the Self-Indication Assumption, SIA \citep{bostrom:2013}, implies that amongst a landscape of possible universes with differing conditions we should expect to emerge in one well-suited for life. For the moment, let us exclude the possibility of destructive infections such as that described in this work (i.e. set $\lambda=0$). In this case, the SIA would favor us living in a crowded universe. However, despite the small volume of the SETI cosmic haystack surveyed thus far \citep{tarter:2010,wright:2018}, a cosmos that maximizes the number of civilizations appears to contradict our observations - the so-called ``Great Silence'' \citep{brin:1983}. Thus, if the SIA holds but infections are impossible, there is an apparent contradiction with existing observations - what one might consider to be an SIA-fueled version of the Fermi paradox. To resolve the paradox, SIA proponents would need to invoke some kind of observer-suppression mechanism, putting pressure on the $\lambda=0$ scenario described above, and thereby providing tacit support for the infection model. Although artificial infection is indeed a candidate mechanism, we caution that it is not a logically unique one. However, it is simultaneously highly effective, economic as a parameterized model, fully physically allowed and represents a plausible extrapolation of humanity's own technology \citep{borgue:2021,ellery:2022}.

As strange as it may seem to try and argue for a non-zero value of a parameter based on the absence of its detection, especially in such a way that depends on anthropic reasoning, we highlight that an anthropic argument like this was used by \citet{weinberg:1987} (and refined in \citealt{martel:1998}) to successfully predict the value of the cosmological constant. Unlike the cosmological constant, $\lambda$ is not a fundamental constant with a known high-energy naturalness problem. The analogy is therefore structural rather than literal: in both cases, a parameter must lie in a small life-permitting region relative to a broad conceivable range.

\subsection{The Infection Fine-Tuning Problem}

One may consider the spawn rate of infections as a basic property of our universe - not a fundamental parameter, like the fine-structure constant or the gravitational constant, but nevertheless an emergent, yet intrinsic, one. Further, this work demonstrates that said parameter must be tuned to an extremely small value, ${\sim}10^{-20}$ per Gyr per star, in order to comport with our mere existence.

This is reminiscent of how the cosmological constant needs to be tuned to $10^{-120}$ times its natural quantum value in order to comport with our existence \citep{weinberg:1989}, and indeed numerous other parameters appear finely-tuned in this way \citep{carr:1979}. Although we stress again that the spawn rate is an emergent parameter, not a fundamental one, it appears to also be enigmatically finely-tuned. And, like those other terms, the problem could be trivially resolved by invoking a landscape (e.g. multiverse) of possibilities and then invoking the weak anthropic principle to explain why our locale appears finely-tuned for life \citep{carter:1974}. Fact A and the related Fermi problem are rarely framed in terms of fine-tuning like this, although \citet{wang:2023} discuss a geometric argument originally made by Denis Sciama concerning the possibility of our universe being on the cusp of inhospitable.

There is another fine-tuning argument made in the field of SETI by \citet{lewis:2025}; one that invokes a Bayesian argument that the cosmos is likely either crowded or empty, both of which are near-saturated and deep states, but unlikely to be in some delicately-balanced, intermediate filling fraction. That same argument seems appropriate here, which would imply $f\sim1/2$ is such a brief, transitory phase that its unlikely we'd live there. But the critical difference is that \citet{lewis:2025} imagined $f\to1$ corresponding to a busy Universe filled with observers, whereas here $f\to1$ would be devoid of them. Accordingly, the weak anthropic principle demands that we don't live in a universe with $f\to1$ when infections are considered. Thus, the \citet{lewis:2025} argument doesn't straight-forwardly apply here, leaving us with the anthropic arguments made earlier in Section~\ref{sec:implications}. As that section showed, subtle differences in the most likely scenario arise depending on what one form of anthropic reasoning one subscribes to. Sadly, a definitive theory of anthropic reasoning does not exist and debates continue to rage, especially over specific Gedankenexperiments like the ``Doomsday Argument''. Nevertheless, we suggest that the cosmological Hart-Tipler problem explored in this work  deserves greater attention to both physicists and philosophers.

\newpage
\appendix
\section{Converting galaxy counts to stars}

A useful alternative normalization is the comoving number density of stars rather than galaxies.  Observational determinations of the local cosmic stellar mass density give values of order
$\rho_{\star,0}\simeq 10^{8.6}\,M_\odot\,{\rm cMpc}^{-3}$, with representative recent estimates including
$$\log_{10}(\rho_{\star}/M_\odot{\rm Mpc}^{-3})=8.59$$ from \citet{li:2009}, $8.78$ from \citet{gallazzi:2008}, and $8.59$ from \citet{moustakas:2013}, as compiled by \citet{madau:2014}. 
For a standard Kroupa/Chabrier-like stellar initial mass function \citep{kroupa:2001,chabrier:2003}, most stars by number are low-mass stars, and a characteristic mean stellar mass of $\langle M_\star\rangle\sim 0.5\,M_\odot$ is therefore adequate for order-of-magnitude purposes.  This gives
\[
n_\star \sim \frac{\rho_{\star,0}}{\langle M_\star\rangle}
      \sim \frac{5\times10^8\,M_\odot\,{\rm cMpc}^{-3}}
                {0.5\,M_\odot}
      \sim 10^9\,{\rm stars}\,{\rm cMpc}^{-3}.
\]
This quantity is less sensitive to the arbitrary lower mass threshold implicit in defining a ``galaxy'' and is arguably a more natural normalization for processes that trace potential stellar systems rather than dark-matter halos or galaxy catalogs.

\newpage
\section{Analytic Half-Filling Estimates in Limiting Cosmologies}
\label{app:half_filling}

In the main text, the infected fraction is written as

\begin{align}
f(t) &= 1 - \exp[-X(t)],
\end{align}

where

\begin{align}
X(t) &= \Xwave(t) + \Xself(t).
\end{align}

The half-filling time is therefore defined implicitly by

\begin{align}
f(t_{1/2}) &= \frac{1}{2},
\end{align}

or equivalently

\begin{align}
X(t_{1/2}) &= \log 2.
\end{align}

In a general $\Lambda$CDM cosmology this condition is most naturally solved numerically. However, useful analytic intuition can be obtained in two limits: a static universe and a matter-dominated universe.

\subsection{Static-universe approximation}
\label{app:static_half_filling}

First consider a static universe, for which

\begin{align}
a(t) &= 1.
\end{align}

If we define $T \equiv (t - \tstart)$ as the elapsed time since the onset of the mutation process. The self-mutation contribution is simply

\begin{align}
\Xself(T) &= \lambda T,
\end{align}

A seed born at time $t'$ has, by time $t$, reached physical and comoving radius

\begin{align}
R(t,t') &= u(t-t'),
\end{align}

with a corresponding volume of

\begin{align}
V(t,t') &= \frac{4\pi}{3} u^3 (t-t')^3.
\end{align}

For a fixed comoving galaxy number density $n_G$, the comoving nucleation rate density is $n_G \lambda$ and thus the wave contribution to the exponent is

\begin{align}
\Xwave(T)
&= \int_{\tstart}^t n_G \lambda \frac{4\pi}{3} u^3 (t-t')^3\,\mathrm{d}t' \nonumber\\
&= \frac{4 \pi n_G \lambda u^3}{3} \int_0^T s^3\,\mathrm{d}s \nonumber\\
&= \frac{  \pi n_G \lambda u^3}{3} T^4.
\end{align}

Therefore, in a static universe,

\begin{align}
X(T) &= \lambda T + \frac{ \pi n_G \lambda u^3 }{ 3 } T^4.
\end{align}

The half-filling time $T_{1/2}$ is the positive solution of

\begin{align}
\lambda T_{1/2} + \frac{ \pi n_G \lambda u^3}{3} T_{1/2}^4 &= \log 2.
\end{align}

Although this quartic equation can be solved exactly, its limiting forms are more transparent. If self-mutation dominates over wave infection (or equivalently $u\to0$), such that $\Xself \gg \Xwave$, then

\begin{align}
T_{1/2} &\simeq \frac{ \log 2 }{ \lambda }.
\end{align}

This is simply the ordinary Poisson half-life for a single galaxy to self-mutate. If, instead, wave infection dominates (i.e. $\Xwave \gg \Xself$), then the quartic becomes

\begin{align}
\frac{\pi n_G \lambda u^3}{3} T_{1/2}^4 &\simeq \log 2,
\end{align}

which has the root

\begin{align}
T_{1/2} &\simeq \Big( \frac{3 \log 2}{\pi n_G \lambda u^3} \Big)^{1/4}.
\end{align}

\subsection{Matter-dominated approximation}
\label{app:matter_half_filling}

A more cosmological analytic limit is obtained by assuming a spatially flat, matter-dominated universe. We write

\begin{align}
a(t) &= \Big( \frac{t}{t_m} \Big)^{2/3},
\end{align}

where, for a universe normalized to $a(t_0)=1$,

\begin{align}
t_m &= \frac{ 2 }{ 3 H _0 \sqrt{\Omega_m} }.
\end{align}

Recall from Equation~(\ref{eqn:Xwave}) that the wave term is

\begin{align}
\Xwave(t) &= \frac{ 4 \pi n_G \lambda u^3}{3}
\int_{\tstart}^t \Big[ \int_{t'}^t \frac{ \mathrm{d}t'' }{ a(t'') } \Big]^3 \mathrm{d}t'.
\end{align}

The inner integral is

\begin{align}
\int_{t'}^t \frac{ \mathrm{d}t'' }{ a(t'') }
&= t_m^{2/3} \int_{t'}^t (t'')^{-2/3}\,\mathrm{d}t'' \nonumber\\
&= 3 t_m^{2/3} \Big( t^{1/3} - (t')^{1/3} \Big).
\end{align}

Therefore

\begin{align}
\Xwave(t)
&= \frac{ 4 \pi n_G \lambda u^3}{3}
\int_{\tstart}^t 27 t_m^2 \Big( t^{1/3}-(t')^{1/3} \Big)^3 \mathrm{d}t' \\
&= 36 \pi n_G \lambda u^3 t_m^2
\int_{\tstart}^t \Big( t^{1/3}-(t')^{1/3} \Big)^3 \mathrm{d}t'.
\end{align}

Expanding the integrand,

\begin{align}
\Big( t^{1/3} - (t')^{1/3} \Big)^3
&= t - 3 t^{2/3} (t')^{1/3} + 3 t^{1/3} (t')^{2/3} - t',
\end{align}

so

\begin{align}
\int_{\tstart}^t \left(t^{1/3}-(t')^{1/3}\right)^3 \mathrm{d}t'
=& t(t-\tstart) - \frac{9}{4}t^{2/3} \left(t^{4/3}-\tstart^{4/3}\right) \nonumber\\
\qquad& + \frac{9}{5} t^{1/3} \left(t^{5/3}-\tstart^{5/3}\right) - \frac{1}{2} \left(t^2-\tstart^2\right).
\end{align}

Thus the matter-dominated wave exponent is

\begin{align}
\Xwave(t) =
36 \pi n_G \lambda u^3 t_m^2
\bigg[
    &t(t-\tstart)
    -
    \frac{9}{4}t^{2/3}
    \left(t^{4/3}-\tstart^{4/3}\right)
    \nonumber\\
    &+
    \frac{9}{5}t^{1/3}
    \left(t^{5/3}-\tstart^{5/3}\right)
    -
    \frac{1}{2}
    \left(t^2-\tstart^2\right)
    \bigg].
\end{align}

The half-filling time is therefore defined by

\begin{align}
\lambda(t_{1/2}-\tstart) + \Xwave(t_{1/2}) &= \log 2.
\end{align}

In general, inverting this requires leaning on numerical roots, but, a particularly simple expression is obtained if the process begins sufficiently early that we may approximate $\tstart=0$. In that case,

\begin{align}
\int_0^t \left(t^{1/3}-(t')^{1/3}\right)^3 \mathrm{d}t
\qquad&= t^2 - \frac{9}{4}t^2 + \frac{9}{5}t^2 - \frac{1}{2}t^2 \nonumber\\
\qquad&= \frac{1}{20}t^2.
\end{align}

Therefore

\begin{align}
X_{\rm wave}(t)
&= 36 \pi n_G \lambda u^3 t_m^2 \left(\frac{1}{20}t^2\right) \nonumber\\
&= \frac{ 9 \pi }{ 5 } n_G \lambda u^3 t_m^2 t^2.
\end{align}

The total exponent becomes

\begin{align}
X(t) &= \lambda t + \frac{ 9 \pi }{ 5 } n_G \lambda u^3 t_m^2 t^2.
\end{align}

The half-filling time satisfies the quadratic equation

\begin{align}
\lambda t_{1/2} + \frac{ 9 \pi }{ 5 } n_G \lambda u^3 t_m^2 t_{1/2}^2 &= \log 2.
\end{align}

Defining

\begin{align}
\beta &= \frac{ 9 \pi }{ 5 } n_G u^3 t_m^2,
\end{align}

this can be written as

\begin{align}
\lambda t_{1/2} + \lambda \beta t_{1/2}^2 &= \log 2,
\end{align}

The positive root is

\begin{align}
t_{1/2} &= \frac{ -1 + \sqrt{ 1 + 4 \beta \log 2/\lambda }
}{
2\beta
}.
\end{align}

In the self-mutation dominated limit, this again reduces to

\begin{align}
t_{1/2} \simeq \frac{\log 2}{\lambda}.
\end{align}

In the wave-dominated matter-era limit,

\begin{align}
\lambda \beta t_{1/2}^2 \simeq \log 2,
\end{align}

and hence

\begin{align}
t_{1/2} \simeq \left( \frac{ 5 \log 2}{ 9\pi n_G \lambda u^3 t_m^2} \right)^{1/2}.
\end{align}

Unlike the static case, the matter-dominated wave term scales as $t^2$, rather than $t^4$, when the process begins near $t=0$. The reason is that in an expanding universe the comoving speed of a fixed physical front is

\begin{align}
\frac{ \mathrm{d}\chi }{ \mathrm{d}t } &= \frac{ u }{ a(t) }.
\end{align}

During matter domination,

\begin{align}
a(t) \propto t^{2/3},
\end{align}

so early fronts acquire a disproportionately large comoving reach. This changes the early-time scaling of the integrated infection volume and hence the dependence of the half-filling time on $u$ and $\lambda$.

\newpage
\section{Comparison to Olson (2015)}
%
Our treatment is mathematically closest to the homogeneous cosmological expansion model of \citet{olson:2015}. In Olson's notation, aggressively expanding civilizations appear with a comoving nucleation-rate density \(f(t)\), and the unsaturated fraction of space is written in the Guth--Tye--Weinberg form
\[
g(t)
=
\exp\left[
-\int_0^t f(t')V(t',t)\,dt'
\right],
\]
where
\[
V(t',t)
=
\frac{4\pi}{3}
\left[
\int_{t'}^t \frac{v\,dt''}{a(t'')}
\right]^3
\]
is the comoving volume reached by an expansion front born at time \(t'\). In the wave-infection part of our model, the corresponding expression is
\[
X_{\rm wave}(t)
=
\int_{t_{\rm start}}^t n_G\lambda_M(t')V_{\rm reach}(t',t)\,dt',
\]
with
\[
V_{\rm reach}(t',t)
=
\frac{4\pi}{3}
\left[
\int_{t'}^t \frac{u\,dt''}{a(t'')}
\right]^3.
\]
Thus the direct notation map is
\[
f(t')_{\rm Olson}
\longleftrightarrow
n_G\lambda_M(t'),
\qquad
v_{\rm Olson}
\longleftrightarrow
u,
\]
and Olson's transformed volume fraction,
\[
1-g(t),
\]
is equivalent to our wave-only infected fraction,
\[
1-\exp[-X_{\rm wave}(t)].
\]

The principal difference is that our model is formulated in terms of discrete susceptible galaxies rather than a continuous transformed volume alone. Each galaxy carries its own spontaneous mutation hazard,
\[
\lambda_M=\tau_M^{-1},
\]
so a target galaxy can become infected in two independent ways: by self-mutation or by being reached by an external infection wave. This leads to the survival probability
\[
P_U(t)
=
\exp[-X_{\rm self}(t)-X_{\rm wave}(t)],
\]
where
\[
X_{\rm self}(t)
=
\int_{t_{\rm start}}^t \lambda_M(t')\,dt'.
\]
This \(X_{\rm self}\) term has no direct analogue in Olson's pure volume-filling expression, because a mathematical point in a continuous volume has no independent self-mutation clock. In our discrete-galaxy interpretation, however, this term is required: it ensures that the model has the correct limit
\[
u\to0
\quad\Rightarrow\quad
f_B(t)=1-\exp[-X_{\rm self}(t)].
\]

A useful way to view the present model is as a deliberately minimal specialization of Olson's framework. We adopt a single expansion speed \(u\), a single susceptible population with fixed comoving number density \(n_G\), a constant or prescribed per-galaxy mutation hazard \(\lambda_M\), zero saturation delay, and no cosmological backreaction from the infected regions. Olson's model is more general: it can accommodate multiple expansion strategies, saturation delays, competition between strategies, resource consumption, and changes to the cosmological background. Those additions are important for many questions, but they are not needed to isolate the basic amplification mechanism emphasized here. The advantage of our formulation is therefore its simplicity: it is essentially the minimal Poisson nucleation-plus-spherical-growth model one can write down for discrete galaxies in an expanding universe.

\newpage
\section*{Acknowledgements}

Special thanks to donors to the Cool Worlds Lab, without whom this kind of research would not be possible:
Steve Larter, Brian Cartmell, Mike Hedlund, Tom Donkin, Bas Gaalen, Emerson Garland, Axel Nimmerjahn, Brad Bueche, Chad Souter, Craig Frederick, Douglas Daughaday, Drew Aron, Ieuan Williams, Jason Rockett, Josh Alley, Mark Elliott, Mathew Farabee, Ryan Provost, Tristan Zajonc, Warren Smith, Guillaume Saint, Marisol Adler, Philip Johnston, Leigh Deacon, Alex Vaal, Andrew Schoen, Benjamin Kingston, Hunter Schiff, Jason Bryant, John Morrison, Jorge Peraza, Michael Simmons, Nicholas Haan, Paul Borisoff, Richard Williams, Stephen Lee, Steven Patterson, Terriss Ford, Trevor Edris \& Zachary Danielson.

\bibliography{manuscript}{}
\bibliographystyle{aasjournalv7_surnames_only}



\end{document}